\documentclass{article}

\usepackage{arxiv}

\usepackage[numbers]{natbib}
\usepackage{amsmath, enumitem}
\usepackage{amsfonts}
\usepackage{upgreek}
\usepackage{graphicx}
\graphicspath{{Figures/}} 
\usepackage{setspace}
\usepackage{multirow}
\usepackage[T1]{fontenc}
\usepackage{fixltx2e}

\title{Bayesian Full-Waveform Inversion with Realistic Priors}

\author{
	Xin Zhang \\
	School of Geosciences \\
	University of Edinburgh\\
	Edinburgh, United Kingdom \\
	\texttt{x.zhang2@ed.ac.uk} \\
	\And
	Andrew Curtis \\
	School of Geosciences \\
	University of Edinburgh\\
	Edinburgh, Unite Kingdom \\
	\texttt{andrew.curtis@ed.ac.uk} \\
}
\begin{document}
\maketitle

\begin{abstract}
Seismic full-waveform inversion (FWI) uses full seismic records to estimate subsurface velocity structure. This requires a highly nonlinear and nonunique inverse problem to be solved, and Bayesian methods have been used to quantify uncertainties in the solution. Variational Bayesian inference uses optimization to provide solutions efficiently. The method has been applied to solve a transmission FWI problem using data generated by known earthquake-like sources, with strong prior information imposed on the velocity. Unfortunately such prior information about velocity structure and earthquake sources is never available in practice. We present the first application of the method in a seismic reflection setting, and with realistically weak prior information. We thus demonstrate that the method can produce high resolution images and reliable uncertainties given practically reasonable prior information.
\end{abstract}

\section{Introduction}
Seismic full waveform inversion (FWI) produces high resolution images of the subsurface directly from seismic waveforms \cite{tarantola1984inversion}. FWI is traditionally solved using optimization by minimizing the difference between predicted and observed seismograms. In such methods a good starting model is often required because of multimodality of the misfit functions caused by the significant nonlinearity of the problem. Those methods also cannot provide accurate estimates of uncertainties, which are required to better understand and interpret the resulting images.

Monte Carlo sampling methods provide a general way to solve nonlinear inverse problems and quantify uncertainties, and have been applied to solve FWI problems \cite{ray2016frequency, zhao2019gradient, gebraad2019bayesian, guo2020bayesian}. However, Monte Carlo methods are usually computationally expensive and all Markov chain Monte Carlo-based methods are difficult to parallelise fully.

Variational inference provides an efficient, fully parallelisable alternative methodology. This is a class of methods that optimize an approximation to a probability distribution describing post-inversion parameter uncertainties \cite{blei2017variational}. The method has been applied to petrophysical inversion \cite{nawaz2018variational, nawaz2019rapid, nawaz2020variational}, travel time tomography \cite{zhang2020seismic},  and more recently to FWI \cite{zhang2020variational}. In the latter study strong prior information is imposed to the velocity structure to limit the space of possible models. Unfortunately such strong information is almost never available in practice. In addition, the method has only been applied to wavefield transmission problems in which seismic data are recorded on a receiver array that lies above the structure to be imaged given known, double-couple (earthquake-like) sources located underneath the same structure. In practice, knowledge of such sources is almost never definitive, and usually depends circularly on the unknown structure itself. In this study, we therefore apply variational inference to solve FWI problems with more practically realistic prior probabilities, and using seismic reflection data acquired from known near-surface sources. 

In the next section we briefly summarise the concept of variational inference, specifically Stein variation gradient descent (SVGD). In section 3 we demonstrate the method by solving an acoustic FWI problem using the Marmousi model with practical prior information. To further explore the method we perform multiple inversions using data from different frequency ranges, and demonstrate that the method can be used with practical prior information to produce high resolution images and uncertainties.

\section{Methods}
\subsection{Stein variational gradient descent (SVGD)}
Bayesian inference solves inverse problems by finding the probability distribution function (pdf) of model $\mathbf{m}$ given prior information and observed data $\mathbf{d}_{obs}$. This is called a \textit{posterior} pdf written $p(\mathbf{m}|\mathbf{d}_{obs})$. By Bayes' theorem,
\begin{equation}
    p(\mathbf{m}|\mathbf{d}_{obs}) = \frac{p(\mathbf{d}_{obs}|\mathbf{m})p(\mathbf{m})}{p(\mathbf{d}_{obs})}
\label{eq:Bayes}
\end{equation}
where $p(\mathbf{m})$ is the \textit{prior} pdf which characterizes the probability distribution of model $\mathbf{m}$ prior to the inversion, $p(\mathbf{d}_{obs}|\mathbf{m})$ is the \textit{likelihood} which represents the probability of observing data $\mathbf{d}_{obs}$ given model $\mathbf{m}$, and $p(\mathbf{d}_{obs})$ is a normalization factor called the \textit{evidence}. 

Variational inference solves Bayesian inference problems using optimization. The method seeks an optimal approximation to the posterior pdf within a predefined family of pdfs, which is achieved by minimizing the Kullback-Leibler (KL) divergence \cite{kullback1951information} between the approximating pdf and the posterior pdf. Variational inference has been shown to be an efficient alternative to Monte Carlo sampling methods for a range of geophysical applications \cite{nawaz2018variational, zhang2020seismic, zhang2020variational}.

Stein variational gradient descent (SVGD) is one such algorithm which iteratively updates a set of models, called particles $\{\mathbf{m}^{i}\}$ generated from an initial distribution $q(\mathbf{m})$ using a smooth transform:
\begin{equation}
T(\mathbf{m}^{i}) = \mathbf{m}^{i} + \epsilon \boldsymbol{\phi} (\mathbf{m}^{i})
\label{eq:transform}
\end{equation}
where $\mathbf{m}^{i}$ is the $i^{th}$ particle, $\boldsymbol{\phi}(\mathbf{m}^{i})$ is a smooth vector function representing the perturbation direction and $\epsilon$ is the magnitude of the perturbation. At each iteration the optimal $\boldsymbol{\phi}$ which produces the steepest direction of KL divergence is found to be:
\begin{equation}
\boldsymbol{\phi} ^{*} (\mathbf{m}) \propto \mathrm{E}_{\{\mathbf{m'} \sim q\}} [\mathcal{A}_{p} k(\mathbf{m'},\mathbf{m})]
\label{eq:phi_qp}
\end{equation}
where $k(\mathbf{m'},\mathbf{m})$ is a kernel function, and $\mathcal{A}_{p}$ is the Stein operator such that for a given smooth function $k(\mathbf{m})$, $\mathcal{A}_{p} k(\mathbf{m}) = \nabla_{\mathbf{m}} \mathrm{log} p(\mathbf{m}) k(\mathbf{m})^{T} + \nabla_{ \mathbf{m} } k( \mathbf{m} )$ \cite{liu2016stein}. The expectation $\mathrm{E}_{\{\mathbf{m'} \sim q\}}$ is calculated using the set of particles $\{\mathbf{m}^{i}\}$, then $\boldsymbol{\phi} ^{*} (\mathbf{m})$ is used to update each particle using equation \ref{eq:transform}. This process is iterated to equilibrium, when the particles are optimally distributed according to the posterior pdf.

In SVGD the choice of kernels can affect the efficiency of the method. In this study we apply a matrix-valued kernel instead of a commonly used scalar kernel to improve efficiency:
\begin{equation}
\mathbf{k}(\mathbf{m'},\mathbf{m}) = \mathbf{Q}^{-1}exp(-\frac{1}{2h}||\mathbf{m}-\mathbf{m'}||^{2}_{\mathbf{Q}})
\end{equation} 
where $\mathbf{Q}$ is a positive definite matrix,  $||\mathbf{m}-\mathbf{m'}||^{2}_{\mathbf{Q}}=(\mathbf{m}-\mathbf{m'})^{T}\mathbf{Q}(\mathbf{m}-\mathbf{m'})$ and $h$ is a scaling parameter. \cite{wang2019stein} showed that by setting $\mathbf{Q}$ to be the Hessian matrix, the method converges faster than with a scalar kernel. However the Hessian matrix is usually expensive to compute. An alternative might be to use the covariance matrix calculated from the particles, but the full covariance matrix may occupy large memory and is difficult to estimate from a small number of samples \cite{ledoit2004well}. We therefore use a diagonal covariance matrix: $\mathbf{Q}^{-1} = diag(\mathrm{var}(\mathbf{m}))$ where $\mathrm{var}(\mathbf{m})$ is the variance estimated from the particles. For those parameters with higher variance, this choice applies higher weights to the posterior gradients  to induce larger perturbations, and also enables interactions with more distant particles.

\subsection{Variational full-waveform inversion}
We apply SVGD to solve an acoustic FWI problem. The wave equation is solved using a time-domain finite difference method. Gradients of the likelihood function with respect to velocity are calculated using the adjoint method \cite{plessix2006review}. For the likelihood function, we assume Gaussian data errors with a diagonal covariance matrix:
\begin{equation}
p(\mathbf{d}_{obs}|\mathbf{m}) \propto exp[-\frac{1}{2}\sum_{i}(\frac{d_{i}^{obs}-d_{i}(\mathbf{m})}{\sigma_{i}})^{2}]
\end{equation}
where $i$ is the index of time samples and $\sigma_{i}$ is the standard deviation of each data point.

\section{Results}
\begin{figure}
\includegraphics[width=1.\linewidth]{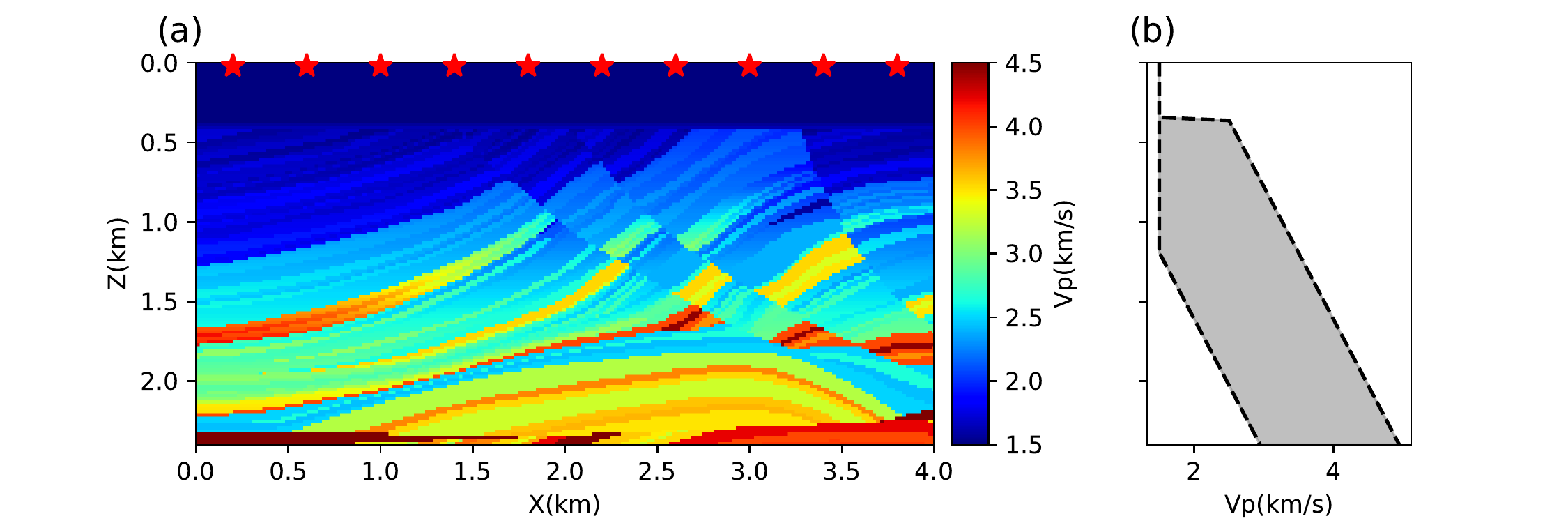}
\caption{\textbf{(a)} The true velocity model. Red stars denote locations of 10 sources. The 200 receivers are equally spaced at 0.36 km depth. \textbf{(b)} The prior distribution of seismic velocity, which is chosen to be a Uniform distribution over an interval of up to 2 km/s at each depth. A lower velocity bound of 1.5 km/s is imposed to ensure the velocity is higher than the acoustic velocity in water.}
\label{fig:true_model}
\end{figure}
We apply the above method to a 2D acoustic full-waveform inversion to recover part of the Marmousi model \cite{martin2006marmousi2} from waveform data (Figure \ref{fig:true_model}). The model is discretised in space using a regular 200 $\times$ 120 grid. Sources are located at 20 m depth in the water layer. 200 equally spaced receivers are located at a depth of 360 meters across the horizontal extent of the model. We generated two waveform datasets using Ricker wavelets with dominant frequency of 4 Hz and 10 Hz respectively. Uncorrelated Gaussian noise with 0.1 standard deviation is added to the data.

\cite{zhang2020variational} and \cite{gebraad2019bayesian} imposed strong prior information (a Uniform distribution over an interval of 0.2 km/s) on the velocity to reduce the complexity of their (identical) inverse problems. Such strong prior information is almost never available in practice. In this study we use ten times weaker prior information: a Uniform distribution over an interval of 2 km/s at each depth (Figure \ref{fig:true_model}b). We also impose an extra lower velocity bound of 1.5 km/s to ensure the rock velocity is higher than the acoustic velocity in water. Velocity in the water layer is fixed to be 1.5 km/s in the inversion. This prior information mimics a practical choice which can be applied in real problems.

\begin{figure}
	\includegraphics[width=1.0\linewidth]{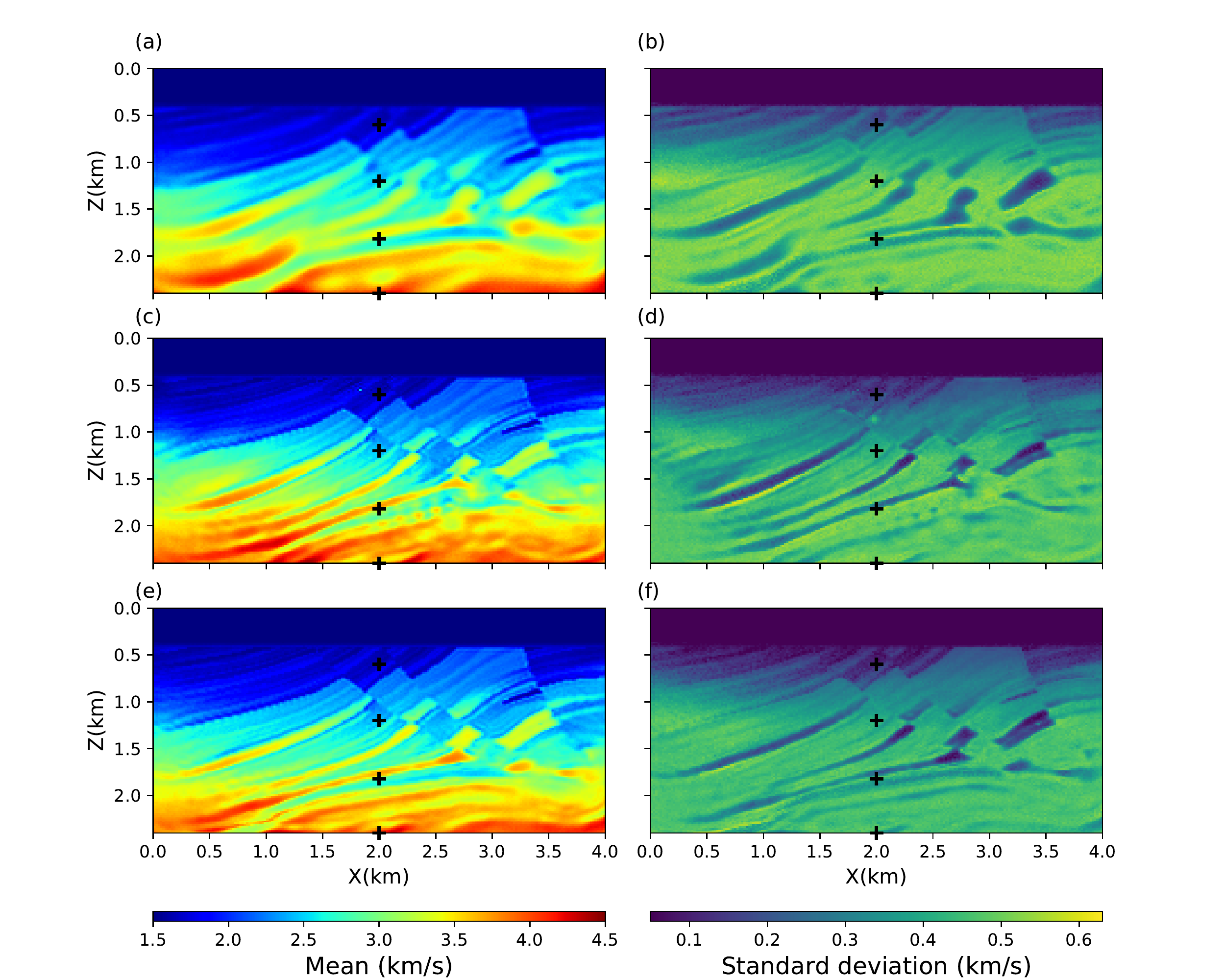}
	\caption{The \textbf{(a), (c)} and \textbf{(e)} mean and \textbf{(b), (d)} and \textbf{(f)} standard deviation models obtained respectively using low frequency data, high frequency data, and using high frequency data but starting from the results of low frequency data. Black pluses denote locations referred to in the main text.}
	\label{fig:meanstd}
\end{figure}

We perform two independent inversions using the two datasets respectively. For each inversion we use 600 particles which are initially generated from the prior distribution and updated using equation \ref{eq:transform} for 600 iterations. Figure \ref{fig:meanstd}a shows the mean model obtained using the low frequency data. In the shallower part ($<$ 2 km) the mean model shows similar features to the true model but has slightly lower resolution than the true model, which probably reveals the resolution limit restricted by the frequency range. In comparison the mean obtained using high frequency data shows higher resolution (Figure \ref{fig:meanstd}c) and is more similar to the true model. In the deeper part ($>$ 2 km) both mean models show differences to the true model: the mean obtained using low frequency data only shows large scale structure, whereas that obtained using high frequency data shows higher resolution details which are different from the true model. This may be because of poor illumination of the deeper part, which causes complex posterior pdfs when using high frequency data and which cannot be represented properly by a small number of particles. However, we also note that the mean model need not reflect the true model in nonlinear problems. Both standard deviation models show features that are related to the mean model. For example, in the shallow part ($<$ 1 km) the standard deviation is lower at locations of lower velocity anomalies, and in the deeper part lower standard deviations are associated with higher velocity anomalies. This phenomenon has also been found by previous studies \cite{gebraad2019bayesian,zhang2020variational}, and probably reflects the fact that waves spend comparatively longer in lower velocity areas resulting in greater sensitivity to those speed parameters.

\begin{figure}
	\includegraphics[width=1.0\linewidth]{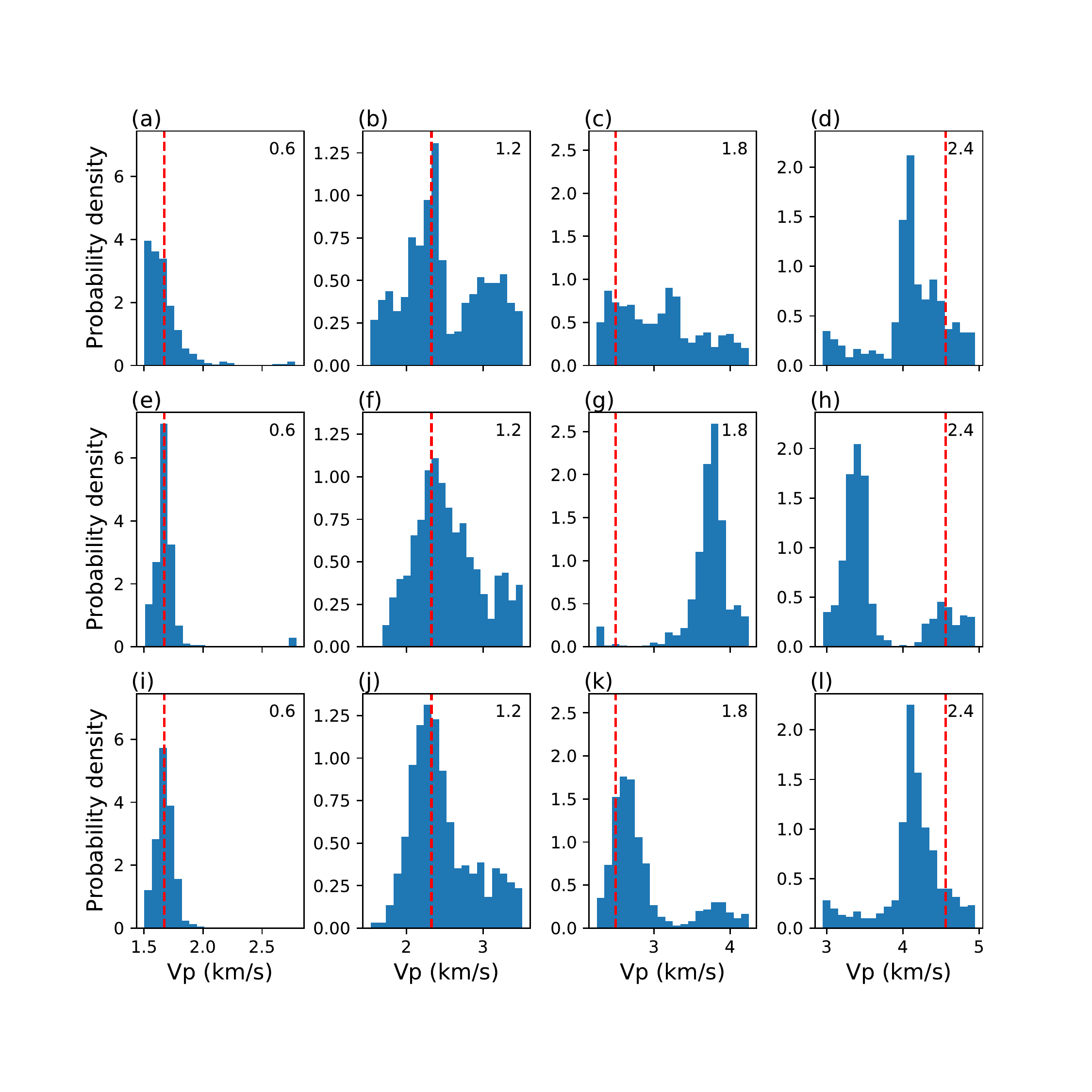}
	\caption{The marginal distributions at horizontal location 2 km and depths of 0.6 km, 1.2 km, 2 km and 2.4 km. The \textbf{top}, \textbf{middle} and \textbf{bottom} rows show marginal distributions obtained using low frequency data, high frequency data only, and using high frequency data but starting from the results of low frequency data, respectively. Dashed red lines show true values.}
	\label{fig:marginals}
\end{figure}
To improve the results in the deeper part, we conducted another inversion by using high frequency data but starting from those particles generated using the low frequency data and run the inversion for 300 iterations. By doing this the mean model shows more similar features to the true model in the deeper part (Figure \ref{fig:meanstd}e). The standard deviation model  (Figure \ref{fig:meanstd}f) also shows smoother structure than the previous results.

To further understand the results, we show marginal velocity distributions at four locations (black pluses in Figure \ref{fig:meanstd}): point (2.0, 0.6) km, (2.0, 1.2) km, (2.0, 1.8) km and (2.0, 2.4) km. Overall the marginal distributions obtained using high frequency data have a tighter distribution. At the shallower points (at depths of 0.6 km and 1.2 km) all the marginal distributions show high probabilities around the true velocity (red lines in Figure \ref{fig:marginals}). At the two deeper points the marginal distributions obtained using low frequency data show high uncertainties due to lower resolution. The marginal distributions obtained using high frequency data only show complex, multimodal distributions, and the high probability area deviates from the true value. In comparison the marginal distributions obtained using the results of low frequency inversion as starting particles show high probabilities around the true value. This clearly indicates that the method can get stuck at local modes in regions of poor illumination when using only high frequency data -- for example, at depth 1.8 km only one incorrect mode is found (Figure \ref{fig:marginals}g). By starting from particles obtained using low frequency data, this issue can largely be resolved.

\section{Discussion}
Since SVGD is based on particles the method can be computationally expensive. For example, the above inversion with 600 iterations took about 6703 CPU hours, which required 74 hours to run using 90 Intel Xeon E5-2630 CPU cores. In practice stochastic minibatch optimization \cite{robbins1951stochastic} can be used to improve the computational efficiency for larger data sets and 3D applications. Since the method does not require good prior information as is required for linearised FWI, the results obtained using a small dataset could be used to provide a reliable starting model for linearised FWI of larger datasets to produce higher resolution models. This study used a diagonal matrix kernel. To improve efficiency of the method other full matrix kernels might be used, for example Hessian matrix kernels \cite{wang2019stein} or Stein variational Newton methods \cite{detommaso2018stein}.

\section{Conclusion}
In this study we presented the first application of variational full-waveform inversion (VFWI) to seismic reflection data. To explore the applicability of the method we imposed realistically weak prior information on seismic velocity: a Uniform prior pdf with 2 km/s interval, and performed multiple inversions using data from different frequency ranges. The results showed that the method can produce high resolution mean and uncertainty models using only high frequency data. However the method can still get stuck at local modes in areas of poor illumination. This can be resolved by using the results obtained from low frequency data to initiate high frequency inversions. We therefore conclude that VFWI may be a useful method to produce high resolution images and reliable uncertainties. 

\section*{Acknowledgments}
The authors thank the Edinburgh Imaging Project sponsors (BP, Schlumberger and Total) for supporting this research. This work has made use of the resources provided by the Edinburgh Compute and Data Facility (ECDF) (http://www.ecdf.ed.ac.uk/).

\bibliographystyle{plainnat}
\bibliography{varfwi}

\appendix

\label{lastpage}

\end{document}